\documentclass[a4paper,12pt]{article}

\usepackage{graphics, epsfig}

\begin{document}

\author{C. Barrab\`es\thanks{E-mail : barrabes@celfi.phys.univ-tours.fr}\\     
\small Laboratoire de Math\'ematiques et Physique Th\'eorique\\
\small  CNRS/UPRES-A 6083, Universit\'e F. Rabelais, 37200 TOURS, France\\
P.A. Hogan\thanks{E-mail : phogan@ollamh.ucd.ie}\\
\small Mathematical Physics Department\\
\small  National University of Ireland Dublin, Belfield, Dublin 4, Ireland}

\title{The Aichelburg--Sexl Boost of an Isolated Source in 
General Relativity}
\date{}
\maketitle

\begin{abstract}
A study of the Aichelburg--Sexl boost of the Schwarzschild 
field is described in which the emphasis is placed on the 
field (curvature tensor) with the metric playing a secondary 
role. This is motivated by a description of the Coulomb 
field of a charged particle viewed by an observer whose 
speed relative to the charge approaches the speed of light. 
Our approach is exemplified by carrying out an Aichelburg--
Sexl type boost on the Weyl vacuum gravitational field due 
to an isolated axially symmetric source. Detailed calculations 
of the boosts transverse and parallel to the symmetry 
axis are given and the results, which differ significantly, 
are discussed.

\end{abstract}
\thispagestyle{empty}
\newpage

\section{Introduction}\indent
It was pointed out many years ago by Bergmann \cite{BER} that to 
a fast--moving observer travelling rectilinearly, the Coulomb 
field of a point charge $e$, with a time--like geodesic world--line 
in Minkowskian space--time, resembles the electromagnetic field 
of a plane electromagnetic wave with a sharpely peaked profile, 
the closer the speed $v$ of the observer relative to the charge 
approaches the speed of light. In fact in the limit $v\rightarrow 1$ 
(we shall use units in which the speed of light $c=1$) the field 
of the charge seen by the observer is that of a plane impulsive 
electromagnetic wave \cite{AE}, \cite{AB}, i.e. a plane electromagnetic wave having a 
Dirac delta function profile. The particle origin of this wave 
is reflected in the fact that the wave front contains a singular 
point. In other words the history of the wave front in Minkowskian 
space--time is a null hyperplane generated by parallel null 
geodesics and on one of these null geodesics the field amplitude 
is singular. The gravitational analogue of this result is contained 
in an influential paper by Aichelburg and Sexl \cite{AS}. They have 
shown that to an observer moving rectilinearly relative to a sphere 
of mass $m$ (the source of the Schwarzschild space--time) with speed 
$v$ the space--time in the limit $v\rightarrow 1$ is a model 
of a plane impulsive gravitational wave. As in the electromagnetic 
case the history of this plane gravitational wave contains a 
null geodesic on which the field amplitude is singular. Many properties 
of this important result have been elucidated \cite{BN} -- \cite{K} 
and it is central 
to the description in general relativity of the high--speed collision 
of black--holes \cite{D}.

The present paper is a study of the Aichelburg--Sexl result, emphasising 
the role of the curvature tensor. In all examples considered the 
Aichelburg--Sexl boost results in a curvature tensor concentrated 
on a null hyperplane in Minkowskian space--time. The singularity in 
the curvature is in the form of a delta function which is 
singular on the null hyperplane. This is consistent with the two 
halves of Minkowskian space-time (one half is the region to the 
future of the null hyperplane and the other half is the region to the 
past of it) being re--attached on the null hyperplane with a matching 
which preserves the intrinsic metric of the hyperplane. The re--attachment 
involves a knowledge of a single function defined on the hyperplane. 
When this function is known the coefficients of the delta function in 
the curvature are calculated from it. The relevant formulas have been derived 
in \cite{BH} and are reproduced in Appendix A here.  Thus for the 
examples given in this paper the relevant matching is explicitly 
calculated. It is then possible to express the metric tensor of the 
re--attached space--time in a coordinate system in which this 
metric tensor is continuous across the null hyperplane \cite{BH}. 
Having established our point of view 
we then  consider what we call the ``Aichelburg--Sexl 
boost" of the Weyl, static, axially symmetric, asymptotically flat 
vacuum gravitational fields. Our results of course depend upon whether 
the observer is moving parallel to the symmetry axis or not. Boosting 
parallel to the symmetry axis reproduces the Aichelburg--Sexl 
space--time in this case while 
boosting transverse to the symmetry axis produces a plane 
impulsive gravitational wave which has an amplitude which is singular 
on one of the null geodesic generators of the history of the plane 
wave front. This singularity however is more severe than in the 
monopole case and has a typical multipole character.

As this paper represents a fresh look at the Aichelburg--Sexl boost 
of the Schwarzschild space--time we want to make it as self--contained 
as possible. To this end and to make clear our point of view the 
paper is organised as follows: the boost of the Coulomb field described 
at the beginning of this introduction is given in detail in 
section 2 and then in section 3 we rederive the Aichelburg--Sexl result 
from our point of view. This prepares the reader for our description 
in section 4 of the boosted Weyl asymptotically flat fields. The paper 
ends with a discussion in section 5. Some fairly extensive 
calculations accompany our work and these are summarised in the 
appendices.

\setcounter{equation}{0}
\section{Coulomb Field}\indent
We begin with the line-element of Minkowskian space--time in 
rectangular Cartesian coordinates and time $\{\bar x, \bar y, \bar z, 
\bar t\}$,
\begin{equation} \label{2.1}
  ds^2=d\bar x^2+d\bar y^2+d\bar z^2-d\bar t^2\ .
\end{equation}
We are using units in which the speed of light $c=1$. Let the 
$\bar t$--axis ($\bar x=\bar y=\bar z=0$) be the time--like 
geodesic world--line of a charge $e$. The Coulomb potential of 
this charge at $(\bar x, \bar y, \bar z, \bar t)$ is given by 
the 1--form
\begin{equation}\label{2.2}
A=-\frac{e}{\bar r}\,d\bar t\ ,\qquad \bar r=(\bar x^2+\bar y^2+\bar z^2
)^{1/2}\ .
\end{equation}
The corresponding electric field is given by the 2--form
\begin{equation}\label{2.3}
F=dA=\frac{e}{\bar r^3}\,\left (\bar x\,d\bar x\wedge d\bar t+
\bar y\,d\bar y\wedge d\bar t+\bar z\,d\bar z\wedge d\bar t\right )\ .
\end{equation}
The field measured by an observer moving in the $-\bar x$ direction 
with speed $v$ relative to the charge is obtained by making in (\ref{2.3}) 
the Lorentz transformation
\begin{equation}\label{2.4}
\bar x=\gamma\,(x-v\,t)\ ,\qquad\bar y=y\ ,\qquad\bar z=z\ ,
\qquad\bar t=\gamma\,(t-v\,x)
\ ,
\end{equation}
with $\gamma =(1-v^2)^{-1/2}$. Thus we can write
\begin{equation}\label{2.5}
F=\frac{e\,\gamma ^{-2}}{R^3}\,\left\{(x-v\,t)\,dx\wedge dt+(y\,dy+
z\,dz)\wedge (dt-v\,dx)\right\}\ ,
\end{equation}
with
\begin{equation}\label{2.6}
R=\left\{(x-v\,t)^2+\gamma ^{-2}(y^2+z^2)\right\}^{1/2}\ .
\end{equation}
We want the limit of (\ref{2.5}) as $v\rightarrow 1$. To 
this end we make the simple observation:
\begin{equation}\label{2.7}
\frac{\gamma ^{-2}}{R^3}=\frac{1}{y^2+z^2}\,\frac{\partial}{\partial x}
\left (\frac{x-v\,t}{R}\right )\ ,
\end{equation}
and so
\begin{equation}\label{2.8}
\lim_{v\longrightarrow 1}\frac{\gamma ^{-2}}{R^3}=(y^2+z^2)^{-1}
\frac{\partial}{\partial x}\left (\frac{x-t}{|x-t|}\right )\ .
\end{equation}
Denoting by $\vartheta (u)$ the Heaviside step function which we 
take to be unity if $u>0$ and zero if $u<0$ we can write
\begin{equation}\label{2.9}
\frac{x-t}{|x-t|}=2\,\vartheta (x-t)-1\ ,
\end{equation}
and so (\ref{2.8}) can be written
\begin{equation}\label{2.10}
\lim_{v\longrightarrow 1}\frac{\gamma ^{-2}}{R^3}=\frac{2\,\delta (x-t)}
{y^2+z^2}\ ,
\end{equation}
with $\delta (x-t)$ the Dirac delta function singular on the null 
hyperplane $x=t$. Remembering that $(x-t)\,\delta (x-t)=0$ we see 
now from (\ref{2.5}) that
\begin{equation}\label{2.11}
\lim_{v\longrightarrow 1}F=\frac{2\,e\,\delta (x-t)}{y^2+z^2}\,
(y\,dy+z\,dz)\wedge (dt-dx)=F_0\ ({\rm say})\ .
\end{equation}
Clearly we can write
\begin{equation}\label{2.12}
F_0=dA_0\ ,\qquad {\rm with}\qquad A_0=e\,\delta (x-t)\,\log (y^2+z^2)\,
(dt-dx)\ ,
\end{equation}
and so $F_0$ is a solution of the source--free Maxwell 
equations which is singular on $x=t$ and also on the null geodesic 
generator of $x=t$ labelled by $y=z=0$. $F_0$ describes an impulsive 
electromagnetic wave (the 2--form $F_0$ is Type N in the Petrov 
classification with degenerate principal null direction given by 
the 1--form $dx-dt$) with profile $\delta (x-t)$.

Substituting the Lorentz transformation (\ref{2.4}) into the potential 
1--form (\ref{2.2}) yields
\begin{equation}\label{2.13}
A=-\frac{e}{R}\,(dt-v\,dx)\ ,
\end{equation}
with $R$ given by (\ref{2.6}). From this we arrive at 
\begin{equation}\label{2.14}
\lim_{v\longrightarrow 1}A=-e\,\frac{(dt-dx)}{|t-x|}\ ,
\end{equation}
which, of course, only makes sense if $x\neq t$. The limit (\ref{2.14}), 
being a pure gauge term for $x>t$ and for $x<t$, is consistent with (\ref{2.11}). 
To see what happens {\it on} $x=t$ in the limit $v\rightarrow 1$ we 
first modify the Coulomb potential (\ref{2.2}) by the addition of a 
gauge term to read (this gauge term is suggested by a clever analogous 
coordinate transformation in the gravitational case \cite{AS})
\begin{equation}\label{2.15}
A=-\frac{e}{\bar r}\,d\bar t-\frac{e\,d\bar x}{\sqrt{\bar x^2+1}}\ .
\end{equation}
This potential 1--form obviously leads to the same electric field 
(\ref{2.3}) 
as that derived from (\ref{2.7}). The Lorentz fransformation (\ref{2.4}) 
applied to (\ref{2.15}) yields
\begin{equation}\label{2.16}
A=-\frac{e}{R}\,(dt-v\,dx)-\frac{e\,(dx-v\,dt)}{\left\{(x-v\,t)^2+
\gamma ^{-2}\right\}^{1/2}}\ .
\end{equation}
Now 
\begin{equation}\label{2.17}
dx-v\,dt=-dt+v\,dx+(1-v)(dx+dt)\ ,
\end{equation}
and so for $v$ near $1$ we can write (\ref{2.16}) as
\begin{equation}\label{2.18}
A=-e\left [\frac{1}{R}-\frac{1}{\left\{(x-v\,t)^2+
\gamma ^{-2}\right\}^{1/2}}\right ]\,(dt-v\,dx)+O\left ((1-v)\right )\ .
\end{equation}
This can further be written
\begin{equation}\label{2.19}
A=-e\,\frac{\partial}{\partial x}\left [\log\left (\frac{x-v\,t+R}{x-v\,t+
\sqrt{(x-v\,t)^2+\gamma ^{-2}}}\right )\right ]\,(dt-v\,dx)
+O\left ((1-v)\right )\ .
\end{equation}
The logarithm term here appeared first in \cite{AS} and it is particularly 
useful when one observes that \cite{AS}
\begin{equation}\label{2.20}
\lim_{v\longrightarrow 1}\log\left (\frac{x-v\,t+R}{x-v\,t+
\sqrt{(x-v\,t)^2+\gamma ^{-2}}}\right )=\left (1-\vartheta (x-t)\right )\,\log 
(y^2+z^2)\ .
\end{equation}
Hence (\ref{2.19}) gives
\begin{equation}\label{2.21}
\lim_{v\longrightarrow 1}A=e\,\frac{\partial}{\partial x}\left (\vartheta 
(x-t)\,\log (y^2+z^2)\right )\,(dt-dx)=e\,\delta (x-t)\,\log (y^2+z^2)\,
(dt-dx)\ ,
\end{equation}
and so we have recovered $A_0$ in (\ref{2.12}). A comprehensive study 
of light--like boosts of electromagnetic multipole fields (which has little overlap in approach with our discussion in section 4 
below of gravitational multipole fields) has been carried out by 
Robinson and R\'ozga \cite{RR}.

We regard the limit of the field (\ref{2.11}) to be the important 
result from which the physical interpretation (that the boosted Coulomb field 
is the field of an impulsive electromagnetic wave in the limit 
$v\rightarrow 1$) follows. The limit of the potential is in this 
sense of secondary importance. Thus in the gravitational case we shall 
place an emphasis on boosting the space--time curvature while the metric 
will play a secondary role.

\setcounter{equation}{0}
\section{Aichelburg--Sexl Boost}\indent
As in \cite{AS} we take the Schwarzschild line--element in 
isotropic coordinates as starting point:
\begin{equation}\label{3.1}
ds^2=\left (1+A\right )^4(d\bar x^2+d\bar y^2+d\bar z^2)-
\frac{\left (1-A\right )^2}{\left (1+A\right )^2}\,d\bar t^2\ ,
\end{equation}
with
\begin{equation}\label{3.2}
A=\frac{m}{2\,\bar r}\ ,\qquad \bar r=\{\bar x^2+\bar y^2+\bar z^2
\}^{1/2}\ .
\end{equation}
The constant $m$ is the mass of the source. Any other asymptotically 
flat form of the line--element in which the coordinates are asymptotically 
rectangular Cartesians and time will do. For example one could start 
with the Kerr--Schild form of (\ref{3.1}) in coordinates $\{\bar x, 
\bar y, \bar z, \bar t\}$ in terms of which the flat background line--element 
is given by (\ref{2.1}). Since we wish to emphasise the Riemann 
curvature tensor we require the curvature tensor components $\bar R_{ijkl}$ 
of the space--time with line--element (\ref{3.1}) in coordinates 
$\{\bar x^i\}=\{\bar x, \bar y, \bar z, \bar t\}$. The non--identically 
vanishing components are
\begin{eqnarray}\label{3.3}
\bar R_{1212} & = & -{m(\bar r^2-3\,\bar z^2)\over\bar r^5}\,
\left (1+A\right )^2\ ,\qquad\bar R_{1313}=-{m(\bar r^2-3\,\bar y^2)\over\bar r^5}\,\left (1+A\right )
^2\ ,\nonumber\\
\bar R_{2323} & = & -{m(\bar r^2-3\,\bar x^2)\over\bar r^5}\,\left (1+A\right )
^2\ ,\qquad\bar R_{1213}=-{3\,m\,\bar y\,\bar z\over\bar r^5}\,
\left (1+A\right )^2\ ,\nonumber\\
\bar R_{1223} & = & {3\,m\,\bar x\,\bar z\over\bar r^5}\,
\left (1+A\right )^2\ ,\qquad
\bar R_{1323}=-{3\,m\,\bar x\,\bar y\over\bar r^5}\,
\left (1+A\right )^2\ ,\nonumber\\
\bar R_{1414} & = & {m\,(\bar r^2-3\,\bar x^2)\over\bar r^5}\,
{\left (1-A\right )^2\over
\left (1+A\right )^4}\ ,\qquad\bar R_{2424}={m\,(\bar r^2-3\,\bar y^2)\over\bar r^5}\,{\left (1-A\right )^2\over
\left (1+A\right )^4}\ ,\nonumber\\
\bar R_{3434} & = & {m\,(\bar r^2-3\,\bar z^2)\over\bar r^5}\,{\left (1-A\right )^2\over
\left (1+A\right )^4}\ ,\qquad\bar R_{1424}=-{3\,m\,\bar x\,\bar y\over\bar 
r^5}\,{\left (1-A\right )^2\over\left (1+A\right )^4}\ ,\nonumber\\
\bar R_{1434} & = & -{3\,m\,\bar x\,\bar z\over\bar 
r^5}\,{\left (1-A\right )^2\over\left (1+A\right )^4}\ ,
\qquad\bar R_{2434}=-{3\,m\,\bar y\,\bar z\over\bar 
r^5}\,{\left (1-A\right )^2\over\left (1+A\right )^4}\ ,
\end{eqnarray}
Now make the Lorentz transformation (\ref{2.4}). 
If $\{x^i\}=\{x, y, z, t\}$ then 
the non--identically vanishing components $R_{ijkl}$ of the Riemann 
tensor in the unbarred coordinates are related to (\ref{3.3}) via
\begin{eqnarray}\label{3.4}
R_{1212} & = & \gamma ^2\left (\bar R_{1212}+v^2\bar R_{2424}\right )\ ,
\qquad R_{1313}=\gamma ^2\left (\bar R_{1313}+v^2\bar R_{3434}\right )\ ,
\nonumber\\
R_{2124} & = & -\gamma ^2v\,\left (\bar R_{2121}+\bar R_{2424}\right )\ ,
\qquad R_{3134}=-\gamma ^2v\,\left (\bar R_{3131}+\bar R_{3434}\right )\ ,
\nonumber\\
R_{3124} & = & -\gamma ^2v\,\left(\bar R_{3121}+\bar R_{3424}\right )\ ,
\qquad R_{1234}=\gamma ^2v\,\left (\bar R_{3121}+\bar R_{3424}\right )\ ,
\nonumber\\
R_{2434} & = & \gamma ^2\left (\bar R_{2434}+v^2\bar R_{1213}\right )\ ,\qquad R_{1213}=\gamma ^2\left (\bar R_{1213}
+v^2\bar R_{2434}\right )\ ,\nonumber\\
R_{1223} & = & \gamma\,\bar R_{1223}\ ,\qquad R_{1323}=\gamma\bar R_{1323}\ ,
\qquad R_{1414}=\bar R_{1414}\ ,\nonumber\\
R_{2424} & = & \gamma ^2\left (\bar R_{2424}+v^2\bar R_{1212}\right )\ ,
\qquad R_{3434}=\gamma ^2\left (\bar R_{3434}+v^2\bar R_{1313}\right )\ ,
\nonumber\\
\qquad R_{1424} & = & \gamma\,\bar R_{1424}\ ,\qquad 
R_{1434}=\gamma\,\bar R_{1434}\ ,\qquad R_{2323}
=\bar R_{2323}\ .
\end{eqnarray}
Of course when (\ref{3.3}) is now substituted into (\ref{3.4}) the barred 
coordinates in (\ref{3.3}) are expressed in terms of the unbarred coordinates 
using (\ref{2.4}). We now make the Aichelburg--Sexl boost of (\ref{3.4}) 
by taking the limit $v\rightarrow 1$. In this limit the rest mass $m\rightarrow 0$ and 
$\gamma\rightarrow\infty$ in such a way that $m\,\gamma =p$ (say) remains finite 
and plays the role of the energy of the source. The calculation of the 
limit is straightforward and makes use of 
\begin{equation}\label{3.5}
\lim_{v\longrightarrow 1}\frac{\gamma ^{-4}}{R^5}=\frac{4}{3}\,
\frac{\delta (x-t)}{(y^2+z^2)^2}\ ,
\end{equation}
which is obtained from (\ref{2.10}) by differentiating with respect 
to $y$ (or $z$) assuming $y, z$ are non--zero. If we write
\begin{equation}\label{3.6}
\tilde R_{ijkl}=\lim_{v\longrightarrow 1}R_{ijkl}, 
\end{equation}
we find that $\tilde R_{ijkl}\equiv 0$ except for
\begin{eqnarray}\label{3.7}
\tilde R_{1212} & = & \tilde R_{2424}=-\tilde R_{1313}=-\tilde R_{3434}=
 \nonumber\\
\tilde R_{3134} & = & -\tilde R_{2124}=-\frac{4\,p\,(y^2-z^2)\,\delta (x-t)}
{(y^2+z^2)^2}\ ,\\
\tilde R_{1213} & = & \tilde R_{2434}=-\tilde R_{3124}=-\tilde R_{2134}
 \nonumber\\
 & = & -\frac{8\,p\,yz\,\delta (x-t)}
{(y^2+z^2)^2}\ ,
\end{eqnarray}
We note that since $R_{jk}=0$ we have $\tilde R_{jk}=
\lim_{v\rightarrow 1}R_{jk}\equiv 0$. Substitution of the Lorentz transformation (\ref{2.4}) into 
the line--element (\ref{3.1}) and then taking the limit $v\rightarrow 1$ 
gives, in agreement with \cite{AS},
\begin{equation}\label{3.9}
\lim_{v\longrightarrow 1}ds^2=dx^2+dy^2+dz^2-dt^2+\frac{4\,p}{|x-t|}
\,(dt-dx)^2\ .
\end{equation}
This holds for $x\neq t$ and is the analogue of (\ref{2.14}). For 
$x>t$ and for $x<t$ this is the line--element of Minkowskian space--
time (which of course is consistent with (\ref{3.7}) and (3.8)). 
For $x>t$ we can write (\ref{3.9}) in the form
\begin{equation}\label{3.10}
ds_+^2=dy_+^2+dz_+^2+2\,du\,dv_+\ ,
\end{equation}
with
\begin{equation}\label{3.11}
y_+=y\ ,\qquad z_+=z\ ,\qquad u=x-t\ ,\qquad v_+=\frac{1}{2}(x+t)+
2p\,\log (x-t)\ .
\end{equation}
For $x<t$ we can write (\ref{3.9}) in the form
\begin{equation}\label{3.12}
ds_-^2=dy_-^2+dz_-^2+2\,du\,dv_-\ ,
\end{equation}
with
\begin{equation}\label{3.13}
y_-=y\ ,\qquad z_-=z\ ,\qquad u=x-t\ ,\qquad v_-=\frac{1}{2}(x+t)-
2p\,\log (t-x)\ .
\end{equation}
>From (\ref{3.10}) and (\ref{3.12}) we see that $u=x-t=0$ is a null 
hyperplane in Minkowskian space--time. The line--elements (\ref{3.10}) 
and (\ref{3.12}) are consistent with having a delta function in the 
Riemann curvature tensor which is singular on $x=t$ provided the 
two halves of Minkowskian space--time, $x>t$ and $x<t$, are attached 
on $x=t$ with \cite{BH}
\begin{equation}\label{3.14}
y_+=y_-\ ,\qquad z_+=z_-\ ,\qquad v_+=F(v_-, y_-, z_-)\ ,
\end{equation}
for some function $F$ defined on $x=t$ for which $\partial F/\partial v_-\neq 0$. 
In the general case , for any such $F$, this will make $x=t$ a model of the most general plane--fronted 
light--like signal propagating through flat space--time with a 
delta function in the curvature tensor singular on $x=t$ and with, in 
general, a delta function in the Ricci tensor aswell (so that the signal 
could be a plane--fronted impulsive gravitational wave accompanied by 
a plane--fronted light--like shell of null matter). The coefficients of 
the delta function in the Riemann tensor and the Einstein tensor are constructed from derivatives 
of the function $F$ in (\ref{3.14}). The explicit formulas from \cite{BH} 
are listed in Appendix A. The line--element of the re--attached 
halves of Minkowskian space--time can be presented, once $F$ is known, 
in a coordinate system in which the metric tensor is continuous across 
the singular hyperplane. This is given in \cite{BH}. Thus the consistency 
of (\ref{3.14}) with the calculated curvature tensor components (\ref{3.7}) 
and (3.8) implies that in the present case (see Appendix A)
\begin{equation}\label{3.15}
v_+=F=v_-+2p\,\log (y^2+z^2)\ ,
\end{equation}
and that the signal in this case with history $x=t$ is an impulsive 
gravitational wave unaccompanied by a light--like shell (because 
$\tilde R_{jk}\equiv 0$). Moreover the function $F$ in (\ref{3.15}) 
is unique up to a (trivial) change of affine parameter $v_+$ 
along the generators of $x=t$. Finally we note from (\ref{3.7}) and 
(3.8) that the amplitudes of the field components (the coefficients 
of the delta function) are singular on the generator $y=z=0$ of the 
null hyperplane $x=t$. This is the remnant of the particle origin of the 
wave described by (\ref{3.7}) and (3.8).

\setcounter{equation}{0}
\section{Boosting an Isolated Source}\indent
To illustrate the usefulness of our approach in which the Riemann 
curvature tensor plays a central role we consider the Weyl, static, axially 
symmetric, asymptotically flat solutions of Einstein's vacuum field 
equations. The line--element is given by
\begin{equation}\label{4.1}
ds^2={\rm e}^{2\sigma -2\psi}(d\bar r^2+\bar r^2d\bar\theta ^2)+{\rm e}^{-2\psi}
\bar r^2\sin ^2\bar\theta\,d\bar\phi^2-{\rm e}^{2\psi}\,d\bar t^2\ ,
\end{equation}
with
\begin{eqnarray}\label{4.2,3}
\psi & = & \sum_{l=0}^{\infty}\frac{A_l}{\bar r^{l+1}}\,P_l(\cos\bar\theta )
\ ,\\
\sigma & = & \sum_{l=0}^{\infty}\,\sum_{m=0}^{\infty}\frac{A_l\,A_m\,(l+1)\,(m+1)}
{(l+m+2)\,\bar r^{l+m+2}}\,\left (P_{l+1}\,P_{m+1}-P_l\,P_m\right )\ .
\end{eqnarray}
Here $P_l(\cos\bar\theta )$ is the Legendre polynomial of degree $l$ 
in the variable $\cos\bar\theta $ and the $A_l$ are constants related to 
the multipole moments of the source. For example \cite{BVM} $A_0=-m\ , 
A_1=-D\ , A_2=-Q-\frac{1}{3}m^3$ where $m$ is the non--zero rest mass 
of the source, $D$ is its dipole moment and $Q$ is its quadrupole moment. 
We can introduce coordinates $\{\bar x, \bar y, 
\bar z\}$ so that $\{\bar x, \bar y, \bar z, \bar t\}$ are asymptotically 
rectangular Cartesians and time simply by putting
\begin{equation}\label{4.4}
\bar x=\bar r\,\sin\bar\theta\,\sin\bar\phi\ ,\qquad \bar y
=\bar r\,\sin\bar\theta\,\cos\bar\phi\ ,\qquad \bar z=\bar r\,\cos\bar\theta\ .
\end{equation}
Then we can write (\ref{4.1}) in the form
\begin{equation}\label{4.5}
ds^2=g_{AB}d\bar x^A\,d\bar x^B+{\rm e}^{2\sigma -2\psi}d\bar z^2-
{\rm e}^{2\psi}d\bar t^2\ ,
\end{equation}
with $\{\bar x^A\}=\{\bar x, \bar y\}$ and
\begin{displaymath}
\left (g_{AB}\right ) = \frac{{\rm e}^{-2\psi}}{(\bar x^2+\bar y^2)}\,\left ( \begin{array}{ccc}
{\rm e}^{2\sigma}\bar x^2+\bar y^2 & 
\left ({\rm e}^{2\sigma}-1\right )\,\bar x\,\bar y \\
\left ({\rm e}^{2\sigma}-1\right )\,\bar x\,
\bar y & {\rm e}^{2\sigma}\bar y^2
+\bar x^2 \end{array} \right )\ ,
\end{displaymath}
and capital letters take values $1, 2$. The Legendre polynomials 
are now functions 
of $\bar z/\bar r$ with $\bar r=\{\bar x^2+\bar y^2+\bar z^2\}^{1/2}$.

We will consider a Lorentz boost in the $-\bar x$ direction given by 
(\ref{2.4}) and also a Lorentz boost in the $-\bar z$ direction (along 
the symmetry axis) given by
\begin{equation}\label{4.6}
\bar x=x\ ,\qquad \bar y=y\ ,\qquad\bar z=\gamma (z-v\,t)\ ,
\qquad\bar t=\gamma (t-v\,z)\ .
\end{equation}
We will then follow this by taking the limit $v\rightarrow 1$ 
(which we are calling the Aichelburg--Sexl boost). In the monopole 
case considered by Aichelburg and Sexl the rest mass $m$ of the 
source was taken to tend to zero in this limit in such a way 
that the energy $p=m\,\gamma$ remained finite. Now the behavior 
of all of the constants $A_l\ (l=0, 1, 2,\dots )$, related to the 
multipole moments of the source, in the limit $v\rightarrow 1$, has 
to be considered. A simple guide is the special case of the Curzon 
\cite{C} solution which is the subcase of 
(\ref{4.1})--(4.3) corresponding to  
two point masses having rest--masses $m_1\ , m_2$ located on 
the $\bar z$--axis at $\bar z=\pm a$ respectively and held in position with a strut. In this case
\begin{equation}\label{4.7}
A_l =-m_1\,a^l-m_2\,(-a)^l\ ,\qquad l=0,1,2,\dots\ .
\end{equation}
In any boost we will assume as before that $m_1, m_2$ tend to 
zero as $v\rightarrow 1$ like $\gamma ^{-1}$, i.e.
\begin{equation}\label{4.8}
m_1=\gamma ^{-1}\hat p_1\ ,\qquad m_2=\gamma ^{-1}\hat p_2\ ,
\end{equation}
with $\hat p_1\ , \hat p_2$ independent of $v$. Thus for a Lorentz 
boost in the $-\bar x$ direction,
\begin{equation}\label{4.9}
A_l=\gamma ^{-1}p_l\ ,\qquad l=0, 1, 2,\dots\ ,
\end{equation}
with
\begin{equation}\label{4.10}
p_l=-\hat p_1\,a^l-\hat p_2\,(-a)^l\ .
\end{equation}
On the other hand for a Lorentz boost in the $-\bar z$ direction 
we assume that $a$ tends to zero as $v\rightarrow 1$ like $\gamma ^{-1}$, 
i.e.
\begin{equation}\label{4.11}
a=\gamma ^{-1}\hat a\ ,
\end{equation}
with $\hat a$ independent of $v$. Hence in this case 
\begin{equation}\label{4.12}
A_l=\gamma ^{-l-1}\,p_l\ ,\qquad l=0, 1, 2,\dots\ ,
\end{equation}
with now
\begin{equation}\label{4.13}
p_l=-\hat p_1\,\hat a ^l-\hat p_2(-\hat a)^l\ .
\end{equation}
The essential difference between (\ref{4.9}) and (\ref{4.12}) is due to the 
fact that (\ref{4.12}) incorporates the Lorentz contraction (\ref{4.11}). 
We shall in general consider only a class of vacuum gravitational 
fields described by (4.1)--(4.3) for which $A_l$ scales with $\gamma$ 
in the form of (\ref{4.9}) for a Lorentz boost in the (transverse) 
$-\bar x$ direction and in the form (\ref{4.12}) for a boost parallel 
to the axis of symmetry (the $-\bar z$ direction). Since we shall be considering later the 
limit $v\rightarrow 1$ and thus $\gamma ^{-1}\rightarrow 0$ we 
can slightly weaken our assumptions by taking (\ref{4.9}) and 
(\ref{4.12}) to be the leading terms in $A_l$ in the two cases for 
small $\gamma ^{-1}$ (thus allowing additional terms which go to 
zero faster than $\gamma ^{-1}$ in the limit $v\rightarrow 1$) without 
affecting the outcome of our calculations. 

We begin by applying the Lorentz transformation (\ref{2.4}) 
in the $-\bar x$ direction to the curvature tensor 
calculated from the line--element (\ref{4.5}). 
The components $\bar R_{ijkl}$ of this curvature tensor, in 
the coordinates $\{\bar x\ , \bar y\ ,\bar z\ ,\bar t\ \}$ are given 
in Appendix B. These are then transformed to $R_{ijkl}$ in the 
coordinates $\{x\ , y\ , z\ , t\}$ introduced in (\ref{2.4}). The 
components $R_{ijkl}$ are given in (B.5). We 
then finally calculate the limit of these components as $v \rightarrow 1$,  
which we denote by
\begin{equation}\label{4.14}
\tilde R_{ijkl}=\lim_{v\longrightarrow 1}R_{ijkl}\ .
\end{equation}
We note that since the space--time described by (4.1)--(4.3) is a 
vacuum space--time, the Ricci tensor $R_{jk}$ vanishes and so $\tilde R_{jk}=
\lim_{v\rightarrow 1}R_{jk}=0$. In carrying out this programme 
of calculations everything hinges on the functions $\psi\ , \sigma$ and 
their first and second derivatives with respect to $\bar x\ , \bar y\ , 
\bar z $ which must 
be evaluated in the coordinates $\{x\ , y\ , z\ , t\}$ for substitution 
into (B.5). With $A_l=\gamma ^{-1}p_l\ ,\ l=0, 1, 2 ,\dots$, as in (\ref{4.9}), and
\begin{equation}\label{4.15}
\bar r=\gamma\,R\ ,\qquad R=\sqrt{(x-v\,t)^2+\gamma ^{-2}(y^2+z^2)}\ ,
\end{equation}
and starting with 
\begin{equation}\label{4.16}
\psi =\sum_{l=0}^{\infty}\frac{A_l}{\bar r^{l+1}}\,P_l\left (
\frac{\bar z}{\bar r}\right )\ ,
\end{equation}
we can write
\begin{equation}\label{4.17}
\psi =\sum_{l=0}^{\infty}p_l\,\frac{\gamma ^{-l-2}}{R^{l+1}}\,P_l(w)\ ,
\end{equation} 
with 
\begin{equation}\label{4.18}
w=\gamma ^{-1}\frac{z}{R}\ .
\end{equation}
It is clear from (\ref{4.17}) that as $v\rightarrow 1$, 
$\psi \rightarrow 0$ like $\gamma ^{-2}$ for $x\neq t$ and 
$\psi \rightarrow 0$ like $\gamma ^{-1}$ for $x=t$. Starting with 
(\ref{4.16}) we obtain 
\begin{eqnarray}\label{4.19}
\frac{\partial\psi}{\partial\bar x} & = & -\sum_{l=0}^{\infty}
p_l\,(x-v\,t)\,\frac{\gamma ^{-l-3}}{R^{l+3}}\,P'_{l+1}\ ,\\
\frac{\partial\psi}{\partial\bar y} & = & -\sum_{l=0}^{\infty}
p_l\,y\,\frac{\gamma ^{-l-4}}{R^{l+3}}\,P'_{l+1}\ ,\\
\frac{\partial\psi}{\partial\bar z} & = & -\sum_{l=0}^{\infty}
p_l\,(l+1)\,\frac{\gamma ^{-l-3}}{R^{l+2}}\,P_{l+1}\ ,
\end{eqnarray}
where the argument of the Legendre polynomials is $w$ given in (\ref{4.18}) 
and the prime denotes differentiation with respect to $w$. We have 
simplified (4.19) and (4.20) using
\begin{equation}\label{4.22}
w\,P'_l+(l+1)\,P_l=P'_{l+1}\ ,
\end{equation}
for $l=0, 1, 2,\dots\  $, and (4.21) has been simplified using 
\begin{equation}\label{4.23}
(w^2-1)\,P'_l=(l+1)\,\left (P_{l+1}-w\,P_l\right )\ ,
\end{equation}
for $l=0, 1, 2,\dots\  $. We need the second derivatives of $\psi$ and 
making use of the relations (\ref{4.22}) and (\ref{4.23}) satisfied 
by the Legendre polynomials we arrive at
\begin{eqnarray}\label{4.24}
\frac{\partial ^2\psi}{\partial\bar x^2} & = & -\sum_{l=0}^{\infty}
p_l\,\frac{\partial}{\partial x}\left (\frac{(x-v\,t)\,\gamma ^{-l-4}
P'_{l+1}}{R^{l+3}}\right )\ ,\\
\frac{\partial ^2\psi}{\partial\bar x\partial\bar y} & = & -
\sum_{l=0}^{\infty}
p_l\,(x-v\,t)\,\frac{\partial}{\partial y}\left (\frac{\gamma ^{-l-3}
P'_{l+1}}{R^{l+3}}\right )\ ,\\
\frac{\partial ^2\psi}{\partial\bar y^2} & = & -
\sum_{l=0}^{\infty}
p_l\,\frac{\partial}{\partial y}\left (\frac{y\,\gamma ^{-l-4}
P'_{l+1}}{R^{l+3}}\right )\ ,\\
\frac{\partial ^2\psi}{\partial\bar z^2} & = & \sum_{l=0}^{\infty}
p_l\,(l+1)\,(l+2)\,\frac{\gamma ^{-l-4}
P_{l+2}}{R^{l+3}}\ ,\\
\frac{\partial ^2\psi}{\partial\bar x\partial\bar z} & = & \sum_{l=0}^{\infty}
p_l\,(l+1)\,(x-v\,t)\,\frac{\gamma ^{-l-4}
P'_{l+2}}{R^{l+4}}\ ,\\
\frac{\partial ^2\psi}{\partial\bar y\partial\bar z} & = & \sum_{l=0}^{\infty}
p_l\,(l+1)\,y\,\frac{\gamma ^{-l-5}
P'_{l+2}}{R^{l+4}}\ .
\end{eqnarray}
To evaluate the limit $v\rightarrow 1$ of quantities involving 
the derivatives (4.19)--(4.21) and (4.24)--(4.29) we make use of 
the following:
\begin{equation}\label{4.30}
\lim_{v\rightarrow 1}\frac{\gamma ^{-l}P_l(w)}{R^{l+1}}=
\frac{(-1)^{l+1}}{l!}\,g^{(l)}\,\delta (x-t)\ ,
\end{equation}
for $l=1, 2, 3, \dots$ , with $g=\log (y^2+z^2)$ and $g^{(l)}=
\partial ^lg/\partial z^l$. This is established by induction 
on $l$. It is true for $l=1$ because
\begin{equation}\label{4.31}
\frac{\gamma ^{-1}P_1(w)}{R^2}=\frac{\gamma ^{-1}w}{R^2}=\frac{\gamma ^{-2}
z}{R^3}\ ,
\end{equation}
and so by (\ref{2.10})
\begin{equation}\label{4.32}
\lim_{v\rightarrow 1}\frac{\gamma ^{-1}P_1(w)}{R^2}=
\frac{2\,z}{y^2+z^2}\,\delta (x-t)=\frac{\partial g}{\partial z}\,\delta (x-t)\ .
\end{equation}
Assume (\ref{4.30}) holds for $l$ and differentiate it with respect 
to $z$ to obtain
\begin{equation}\label{4.33}
\lim_{v\rightarrow 1}\frac{\gamma ^{-l-1}}{R^{l+2}}\left\{-(l+1)\,w\,P_l
+(1-w^2)\,P'_l\right\}=\frac{(-1)^{l+1}}{l!}g^{(l+1)}\,\delta (x-t)\ .
\end{equation}
Using (\ref{4.23}) this can be simplified to read
\begin{equation}\label{4.34}
\lim_{v\rightarrow 1}\frac{\gamma ^{-l-1}P_{l+1}(w)}{R^{l+2}}
=\frac{(-1)^{l+2}}{(l+1)!}g^{(l+1)}\,\delta (x-t)\ ,
\end{equation}
and (\ref{4.30}) holds for $(l+1)$ if it holds for $l$. 

In addition to (\ref{4.30}) we shall require one more result:
\begin{equation}\label{4.35}
\lim_{v\rightarrow 1}\frac{\gamma ^{-l-2}P'_{l+1}(w)}{R^{l+3}}
=\frac{2\,(-1)^l}{l!}\,h^{(l)}\,\delta (x-t)\ ,
\end{equation}
for $l=0, 1, 2, \dots $ , with $h=(y^2+z^2)^{-1}$ and $h^{(l)}=
\partial ^lh/\partial z^l$. This can be 
established by induction on $l$ or more directly by differentiating 
(\ref{4.30}) with respect to $y$ to obtain
\begin{equation}\label{4.36}
-\lim_{v\rightarrow 1}\frac{\gamma ^{-l-2}y}{R^{l+3}}\,\left\{
(l+1)\,P_l+w\,P'_l\right \}=\frac{(-1)^{l+1}}{l!}\,
\frac{\partial}{\partial y}\left (g^{(l)}\right )\,\delta (x-t)\ ,
\end{equation}
for $l=1, 2, 3,\dots $ . If we use (\ref{4.22}) in the left hand side 
here and in the right hand side use
\begin{equation}\label{4.37}
\frac{\partial}{\partial y}\left (g^{(l)}\right )=\frac{\partial ^l}
{\partial z^l}\left (\frac{\partial g}{\partial y}\right )=
2\,y\,\frac{\partial ^l}{\partial z^l}(y^2+z^2)^{-1}=2\,y\,h^{(l)}\ ,
\end{equation}
then, assuming $y\neq 0$, (\ref{4.36}) becomes (\ref{4.35}) 
for $l=1, 2, 3, \dots\ $. That (\ref{4.35}) also holds for 
$l=0$ follows from
\begin{equation}\label{4.38}
\lim_{v\rightarrow 1}\frac{\gamma ^{-2}P'_1(w)}{R^3}
=\lim_{v\rightarrow 1}\frac{\gamma ^{-2}}{R^3}=2\,h\,\delta (x-t)\ ,
\end{equation}
with the last equality coming from (\ref{2.10}).

With the help of (\ref{4.30}) and (\ref{4.35}) limits of 
the derivatives (4.19)--(4.21) and (4.24)--(4.29) can be 
evaluated which are required for the calculation of 
$\tilde R_{ijkl}$ in (\ref{4.14}). The reader can easily 
evaluate the limits needed with the information given above. We 
shall mention here, as a guide, some of the required limits. 
For example using (4.27) and (\ref{4.30}) we obtain
\begin{equation}\label{4.39}
\lim_{v\rightarrow 1}\gamma ^2\frac{\partial ^2\psi}{\partial\bar z^2}
=\sum_{l=0}^{\infty}p_l\,\frac{(-1)^{l+1}}{l!}g^{(l+2)}\,
\delta (x-t)\ .
\end{equation}
By (4.28) and (\ref{4.35}),
\begin{equation}\label{4.40}
\lim_{v\rightarrow 1}\frac{\partial ^2\psi}{\partial\bar x\partial\bar z}
=0\ ,
\end{equation}
and we have here made use of $(x-t)\,\delta (x-t)=0$. Using 
(4.29) and (\ref{4.35}) we find that
\begin{equation}\label{4.41}
\lim_{v\rightarrow 1}\gamma ^2\frac{\partial ^2\psi}{\partial\bar y\partial
\bar z}=\sum_{l=0}^{\infty}p_l\,\frac{2\,(-1)^{l+1}}{l!}\,
y\,h^{(l+1)}\,\delta (x-t)\ .
\end{equation}
We can write
\begin{equation}\label{4.42}
2\,y\,h^{(l+1)}=\frac{\partial ^{l+1}}{\partial z^{l+1}}
\left (\frac{2\,y}{y^2+z^2}\right )=\frac{\partial ^{l+1}}{
\partial z^{l+1}}\left (\frac{\partial g}{\partial y}\right )
=\frac{\partial ^2}{\partial z\partial y}\,g^{(l)}\ ,
\end{equation}
and thus (\ref{4.41}) becomes
\begin{equation}\label{4.43}
\lim_{v\rightarrow 1}\gamma ^2\frac{\partial ^2\psi}{
\partial\bar y\partial\bar z}=\sum_{l=0}^{\infty}
p_l\,\frac{(-1)^{l+1}}{l!}\,\frac{\partial ^2g^{(l)}}{\partial z\partial y}
\,\delta (x-t)\ .
\end{equation}
Similarly from (4.24), (4.25) using (4.35) and $(x-t)\,\delta (x-t)=0$ 
we find that
\begin{equation}\label{4.44}
\lim_{v\rightarrow 1}\gamma ^2\frac{\partial ^2\psi}{\partial\bar x^2}=0=
\lim_{v\rightarrow 1}\gamma\,\frac{\partial ^2\psi}{\partial\bar x\partial
\bar y}\ .
\end{equation}
Finally by (4.26) using (\ref{4.35}), 
\begin{equation}\label{4.45}
\lim_{v\rightarrow 1}\gamma ^2\frac{\partial ^2\psi}{\partial\bar y^2}
=\sum_{l=0}^{\infty}p_l\,\frac{2\,(-1)^{l+1}}{l!}\,\frac{\partial}{\partial 
y}\left (y\,h^{(l)}\right )\,\delta (x-t)\ .
\end{equation}
But
\begin{equation}\label{4.46}
\frac{\partial}{\partial y}\left (2\,y\,h^{(l)}\right )=
\frac{\partial ^{l+1}}{\partial y\partial z^l}\left (\frac{2\,y}{
y^2+z^2}\right )=\frac{\partial ^2g^{(l)}}
{\partial y^2}\ ,
\end{equation}
and so (\ref{4.45}) reads 
\begin{equation}\label{4.47}
\lim_{v\rightarrow 1}\gamma ^2\frac{\partial ^2\psi}{\partial\bar y ^2}
=\sum_{l=0}^{\infty}p_l\,\frac{(-1)^{l+1}}{l!}\,\frac{\partial ^2g^{(l)}}
{\partial y^2}\,\delta (x-t)\ .
\end{equation}

There remains to be considered the function $\sigma$ given by (4.3). 
It is clear from (4.3) that $\sigma$ is a linear combination with 
constant coefficients of $\psi ^2$--terms. Hence as $v\rightarrow 1$ we 
have $\sigma\rightarrow 0$ like $\gamma ^{-4}$ for $x\neq t$ (compared 
with $\gamma ^{-2}$ for $\psi$) and $\sigma\rightarrow 0$ like $\gamma ^{-2}$ 
for $x=t$ (compared with $\gamma ^{-1}$ for $\psi$). In addition 
all first and second derivatives of $\sigma$ {\it vanish} in the limit 
$v\rightarrow 1$, including the derivatives multiplied by $\gamma$ or 
$\gamma ^2$ (such as the left hand sides of (4.39)--(4.41) and 
(4.43)--(4.45) with $\psi$ replaced by $\sigma$) which are required 
for the calculation of $\tilde R_{ijkl}$ in (4.14). We are now ready to calculate 
$\tilde R_{ijkl}$. We first find that the following components 
are non--zero:
\begin{eqnarray}\label{4.48}
\tilde R_{2434} & = & \tilde R_{2131}=-\tilde R_{2134}=-\tilde R_{3124}=
2\,\lim_{v\rightarrow 1}\gamma ^2\frac{\partial ^2\psi}{\partial\bar z
\partial\bar y}\ ,\nonumber\\
{} & = & 2\,\sum_{l=0}^{\infty}p_l\,\frac{(-1)^{l+1}}{l!}\,
\frac{\partial ^2g^{(l)}}{\partial y\partial z}\,\delta (x-t)\ ,
\end{eqnarray}
by (\ref{4.43}). Next we find
\begin{eqnarray}\label{4.49}
\tilde R_{2424} & = & \tilde R_{2121}=-\tilde R_{2124}=
\lim_{v\rightarrow 1}\gamma ^2\left (2\,\frac{\partial ^2\psi}
{\partial\bar y^2}+\frac{\partial ^2\psi}{\partial\bar x^2}\right )\ ,\nonumber\\
{} & = & 2\,\sum_{l=0}^{\infty}p_l\,\frac{(-1)^{l+1}}{l!}\,
\frac{\partial ^2g^{(l)}}{\partial y^2}\,\delta (x-t)\ ,
\end{eqnarray}
by (\ref{4.44}) and (\ref{4.47}). And finally we obtain
\begin{eqnarray}\label{4.50}
\tilde R_{3434} & = & \tilde R_{3131}=-\tilde R_{3134}=
\lim_{v\rightarrow 1}\gamma ^2\left (2\,\frac{\partial ^2\psi}
{\partial\bar z^2}+\frac{\partial ^2\psi}{\partial\bar x^2}\right )\ ,\nonumber\\
{} & = & 2\,\sum_{l=0}^{\infty}p_l\,\frac{(-1)^{l+1}}{l!}\,
\frac{\partial ^2g^{(l)}}{\partial z^2}\,\delta (x-t)\ ,
\end{eqnarray}
by (\ref{4.39}) and (\ref{4.44}). Noting that $g=\log (y^2+z^2)$ is 
a harmonic function for $y^2+z^2\neq 0$ so that
\begin{equation}\label{4.51}
\frac{\partial ^2g}{\partial y^2}+\frac{\partial ^2g}{\partial z^2}=0\ ,
\end{equation}
the right hand side of (\ref{4.50}) is minus the right hand 
side of (\ref{4.49}). We have now arrived at the analogue of (\ref{3.7}) 
and (3.8) in the monopole case and we see that the singularity 
in (4.48)--(4.50) on $y=z=0$ is more severe than in the monopole case, 
involving derivatives of $g$ greater than the second. 

Taking the limit $v\rightarrow 1$ of the line--element (\ref{4.5}) following 
the coordinate transformation (\ref{2.4}) results in 
\begin{equation}\label{4.52}
\lim_{v\rightarrow 1}ds^2=dx^2+dy^2+dz^2-dt^2-\frac{4\,p_0}{|x-t|}
\,(dt-dx)^2\ ,
\end{equation}
for $x\neq t$ (we note that $p_0=-p$ in (\ref{3.9})). We note that 
only $p_0$ appears in (\ref{4.52}) and so (\ref{4.52}) inherits 
information only from the monopole structure of the source. The 
curvature tensor (\ref{4.48})--(\ref{4.50}) is influenced by all 
the multipole moments of the source and thus contains more information 
than (\ref{4.52}). The discussion 
following (\ref{3.9}) applies again here. The hyperplane $x=t$ is 
null and the Riemann tensor $\tilde R_{ijkl}$ given by (4.48)--(4.50) 
is consistent with re--attaching the two halves of Minkowskian space--
time $x>t$ and $x<t$ with the matching (\ref{3.14}) provided the 
function $F$ in (\ref{3.14}) is correctly chosen. The formulas for 
determining $F$ are obtained by equating to zero the components (A.1)--
(A.4) of the surface stress--energy tensor (which as before is zero 
because $\tilde R_{jk}=0$; the signal with history $x=t$ is a pure 
impulsive gravitational wave) and by equating the expressions for 
$\tilde R_{ijkl}$ in (A.5) and (A.6) with those given above in 
(4.48)--(4.50) (in the course of which we can put $y_-=y\ , z_-=z$). The 
result is that now (\ref{3.15}) is generalised to 
\begin{equation}\label{4.53}
v_+=F=v_--2\,\sum_{l=0}^{\infty}p_l\,\frac{(-1)^{l}}{l!}\,
\frac{\partial ^lg}{\partial z^l}\ ,
\end{equation}
with $g=\log (y^2+z^2)$. This agrees with (\ref{3.15}) in the special 
monopole case when $p_0=-p\ , p_l=0$ for $l=1, 2, 3,\dots $.

We now consider an Aichelburg--Sexl boost of (4.1)--(4.3) in the 
$-\bar z$ direction. We begin again with the curvature tensor $\bar R_{ijkl}$ 
given in Appendix B and transform to the coordinates $\{x, y, z, t\}$ 
using the Lorentz boost (\ref{4.6}) (instead of (\ref{2.4})). In 
this case starting with (\ref{4.16}) with $A_l=\gamma ^{-l-1}p_l$ 
$(l=0, 1, 2, \dots\ )$, as in (\ref{4.12}),
\begin{equation}\label{4.54}
\psi =\sum_{l=0}^{\infty}p_l\,\frac{\gamma ^{-2l-2}}{\hat R^{l+1}}\,
P_l(\hat w)\ ,
\end{equation}
where
\begin{equation}\label{4.55}
\bar r=\gamma\,\hat R\ ,\qquad \hat R=\sqrt{\gamma ^{-2}(x^2+y^2)
+(z-v\,t)^2}\ ,
\end{equation}
and
\begin{equation}\label{4.56}
\hat w=\frac{\bar z}{\bar r}=\frac{z-v\,t}{\hat R}\ .
\end{equation}
Once again we must calculate the first and second derivatives 
of $\psi\ , \sigma$ with respect to $\{\bar x, \bar y, \bar z \}$ 
and express them in terms of $\{x, y, z, t\}$ via (\ref{4.6}) for 
substitution into the curvature components $R_{ijkl}$ listed in 
(B.7). Then the limit $v\rightarrow 1$ of these components is taken 
to arrive at $\tilde R_{ijkl}$ in this case. The calculation parallels 
the one above but with some significant differences. To keep this 
section within reasonable bounds we have given these calculations 
in Appendix C. The result is that $\tilde R_{ijkl}$ given in (C.20) 
and (C.21) is the same curvature tensor as (3.7) and (3.8). The 
presence of the higher multipole moments than the monopole is 
lost in this boost. This is not surprising because the scaling 
of $A_l$ in ({\ref{4.12}) reflects the Lorentz contraction of the 
source in the axial direction. The shape of the source remains the same 
but scaled down to the boosted observer \cite{R}. As $v\rightarrow 1$ 
it appears increasingly as a monopole source and hence we recover 
the original Aichelburg--Sexl result in this case.

\setcounter{equation}{0}
\section{Discussion}\indent
We have been motivated in our approach by the light--like boost 
of the Coulomb field described in section 2. The main conclusion 
we have drawn from this is that the physical significance of the 
boost is deduced from the behavior of the electric field of the 
charged particle. This has led us to place the emphasis in the 
gravitational case on the gravitational field described by the 
Riemann curvature tensor. Incidentally the difficulties inherent in 
extending the light--like boost to the Reissner--Nordstrom case 
have been discussed very clearly in \cite{AE}. The metric plays 
a secondary role for us because once we have recognised that the 
limit is a space--time which is flat to the future and to the past 
of a null hyperplane and has a curvature tensor with a delta function 
singularity on the null hyperplane, then we can reconstruct geometrically 
this curvature tensor by attaching the two halves of flat space--time, 
on the null hyperplane, with an appropriate matching. In addition the 
curvature tensor contains more information than the metric. For example 
in the case of the boosted multipole field described in section 4, 
the limit (\ref{4.52}) of the metric tensor involves only the constant 
$p_0$ which is a remnant of the monopole part of the original 
multipole field. The boosted curvature tensor (\ref{4.48})--(\ref{4.50}) 
in contrast involves all of the constants $p_l\ ,\ l\ge 0$ and thus 
inherits a contribution from all of the original multipole moments.

In the axially symmetric case discussed in detail in section 4, with 
further details in Appendices B and C, there is, as one would 
expect, a significant difference in the outcome if one considers 
a light--like boost transverse to the symmetry axis or parallel 
to the symmetry axis. The results however are consistent with what 
one would expect: (a) Such a boost parallel to the symmetry axis 
yields a result identical to the spherically symmetric monopole 
case, due to Lorentz contraction of the source and (b) the light--like boost 
transverse to the symmetry axis yields a plane impulsive gravitational 
wave which has a singularity along a null geodesic generator of the 
history of the wave which is more severe than in the monopole case 
(as can be explicitly seen by comparing (4.48)--(4.50) with (3.7) 
and (3.8)). The space--time with curvature tensor (4.48)--(4.50) 
can now be used to study the non--axially symmetric high speed 
collision of two sources of the vacuum Weyl gravitational field 
(4.1)--(4.3). The axially symmetric high speed collision of two such 
sources is indistinguishable from the high speed collision of two 
spherically symmetric sources (two black--holes) discussed in \cite{D}. 

\noindent
\section*{Acknowledgment}\noindent
This collaboration has been funded by the Minist\`ere des Affaires 
\'Etrang\`eres, D.C.R.I. 220/SUR/R.

\vskip 8truepc

\appendix
\setcounter{equation}{0}
\section{Construction of Plane--Fronted Light--Like Signal}\indent
When the two halves of Minkowskian space--time $x>t$ and $x<t$ with 
line--elements (\ref{3.10}) and (\ref{3.12}) are attached on the 
null hyperplane $x=t\ (u=0)$ with the matching (\ref{3.14}) there is 
in general a $\delta (x-t)$--term in the curvature tensor of the 
re--attached space--times and also in the Einstein tensor \cite{BH}. The coefficients 
of the delta function can be calculated in terms of drivatives of 
the function $F$ in (\ref{3.14}) using the technique of Barrab\`es 
and Israel \cite{BI}. The coefficients of the delta function in the 
Einstein tensor of the re--attached space--times have the form 
$16\pi\,S_{ij}$ where $S_{ij}$ are the components of the surface stress--
energy tensor of the light--like shell. In the case of (\ref{3.10}) 
and (\ref{3.12}) with the matching (\ref{3.14}) the non--vanishing 
components of $S_{ij}$ in the coordinates $\{x_-, y_-, z_-, t_-\}$ 
(with $x_--t_-=u\ ,\ x_-+t_-=2v_-$) are given by
\begin{eqnarray}\label{A1}
16\pi\,S_{11} & = & 16\pi\,S_{44}=-16\pi\,S_{14}=-\frac{2}{F_{v_-}}\,
\left (F_{y_-y_-}+F_{z_-z_-}\right )\ ,\\
16\pi\,S_{12} & = & -16\pi\,S_{24}=-\frac{2\,F_{y_-v_-}}{F_{v_-}}\ ,\\
16\pi\,S_{13} & = & -16\pi\,S_{34}=-\frac{2\,F_{z_-v_-}}{F_{v_-}}\ ,\\ 
16\pi\,S_{22} & = & 16\pi\,S_{33}=\frac{2\,F_{v_-v_-}}{F_{v_-}}\ .
\end{eqnarray}
The subscripts on $F$ indicate partial derivatives and $F_{v_-}\neq 0$. 
If the components $S_{ij}$ vanish, as they must for the examples 
given in sections 3 and 4 above, then the Riemann tensor of the re--attached 
space--times has the non--identically vanishing components
\begin{eqnarray}\label{A2}
\tilde R_{1212} & = & \tilde R_{2424}=-\tilde R_{1313}=-\tilde R_{3434}=
 \nonumber\\
\tilde R_{3134} & = & -\tilde R_{2124}=\frac{F_{y_-y_-}}{F_{v_-}}
\,\delta (x-t)\ ,\\
\tilde R_{1213} & = & \tilde R_{2434}=-\tilde R_{3124}=-\tilde R_{2134}
 \nonumber\\
 & = & \frac{F_{y_-z_-}}{F_{v_-}}\,\delta (x-t)\ .
\end{eqnarray}
When applying these formulas, for example in section 3, we can put 
$y_-=y\ , z_-=z$ and the equations to be satisfied by $F(v_-, y, z)$ 
are
\begin{equation}\label{A3}
F_{yy}+F_{zz}=0\ ,\qquad F_{yv_-}=0\ ,\qquad F_{zv_-}=0\ ,
\qquad F_{v_-v_-}=0\ ,
\end{equation}
and 
\begin{equation}\label{A4}
\frac{F_{yy}}{F_{v_-}}=-\frac{4p\,(y^2-z^2)}{(y^2+z^2)^2}
\ ,\qquad \frac{F_{yz}}{F_{v_-}}=-\frac{8p\,yz}{(y^2+z^2)^2}\ ,
\end{equation}
The general solution is easily found to be
\begin{equation}\label{A5}
v_+=F(v_-, y, z)=v_-+2p\,\log (y^2+z^2)\ ,
\end{equation}
modulo a (trivial) change of affine parameter $v_+\rightarrow 
a_1\,v_++a_2\,y+a_3\,z+a_4$ where $a_1, a_2, a_3, a_4$ are real constants. 
The continuous form of the metric for a general plane--fronted 
light--like signal is given in \cite{BH}.

\vskip 4truepc
\setcounter{equation}{0}
\section{Curvature Tensor of the Weyl Space--Times}\indent
Let $\bar R_{ijkl}$ denote the curvature tensor components for 
the space--time with line--element (\ref{4.5}) in the 
coordinates $\{\bar x^i\}=\{\bar x^A\ , \bar x^3=\bar z\ , \bar 
x^4=\bar t\}$ with $A=1, 2$. The non--identically vanishing 
components in a convenient form for our purposes are
\begin{eqnarray}\label{B1}
\bar R_{ABCD} & = & {}^{(2)}\bar R_{ABCD}+\frac{1}{4}{\rm e}^
{2\psi -2\sigma}\left (\frac{\partial g_{AD}}{\partial\bar z}\,
\frac{\partial g_{BC}}{\partial\bar z}- \frac{\partial g_{AC}}
{\partial\bar z}\,\frac{\partial g_{BD}}{\partial\bar z}\right )\ ,\nonumber\\
\bar R_{ABC3} & = & \frac{1}{2}\,\frac{\partial}{\partial\bar z}
\left (\frac{\partial g_{BC}}{\partial\bar x^A}-
\frac{\partial g_{AC}}{\partial\bar x^B}\right )+ \frac{1}{2}\,{\rm e}^{2\psi -2\sigma}\,\left (
\frac{\partial g_{33}}{\partial\bar x^B}\,\frac{\partial g_{AC}}
{\partial\bar z}-\frac{\partial g_{33}}{\partial\bar x^A}\,\frac{\partial g_{BC}}
{\partial\bar z}\right )  \nonumber\\
{} & {} & +\frac{1}{2}\,
g^{EF}\,\left (\frac{\partial g_{AE}}{\partial\bar z}\,
[BC,F]-\frac{\partial g_{BF}}{\partial\bar z}\,[AC,E]\right )
\ ,\nonumber\\
\bar R_{3434} & = & -\frac{1}{2}\frac{\partial ^2g_{44}}{\partial
\bar z^2}-\frac{1}{4}g^{EF}\,\frac{\partial g_{33}}{\partial\bar x^E}\,
\frac{\partial g_{44}}{\partial\bar x^F}+\frac{1}{4}{\rm e}^{2\psi -2\sigma}
\frac{\partial g_{33}}{\partial\bar z}\,\frac{\partial g_{44}}
{\partial\bar z} \nonumber\\
{} & {} & -\frac{1}{4}{\rm e}^{-2\psi}\,\left (\frac{\partial g_{44}}
{\partial\bar z}\right )^2\ ,\nonumber\\
\bar R_{A3B3} & = & -\frac{1}{2}\left (\frac{\partial ^2g_{AB}}{\partial 
\bar z^2}+\frac{\partial ^2g_{33}}{\partial\bar x^A\partial\bar x^B}
\right )+\frac{1}{4}g^{EF}\,\frac{\partial g_{AE}}{\partial\bar z}\,
\frac{\partial g_{BF}}{\partial\bar z}+\frac{1}{2}g^{EF}\,
[AB,E]\,\frac{\partial g_{33}}{\partial\bar x^F}  \nonumber\\
{} & {} & +\frac{1}{4}{\rm e}^{2\psi -2\sigma}\left (\frac{\partial g_{33}}
{\partial\bar x^A}\,\frac{\partial g_{33}}{\partial\bar x^B}+
\frac{\partial g_{AB}}{\partial\bar z}\,\frac{\partial g_{33}}{\partial\bar z}\right )\ ,\nonumber\\
\bar R_{A434} & = & -\frac{1}{2}\frac{\partial ^2g_{44}}{\partial\bar z
\partial\bar x^A}+\frac{1}{4}g^{EF}\,\frac{\partial g_{AE}}{\partial\bar z}
\,\frac{\partial g_{44}}{\partial\bar x^F}+
\frac{1}{4}{\rm e}^{2\psi -2\sigma}
\frac{\partial g_{33}}{\partial\bar x^A}\,
\frac{\partial g_{44}}{\partial\bar z} \nonumber\\
{} & {} & -\frac{1}{4}{\rm e}^{-2\psi}\,\frac{\partial g_{44}}{\partial
\bar x^A}\,\frac{\partial g_{44}}{\partial\bar z}\ ,\nonumber\\
\bar R_{A4B4} & = & -\frac{1}{2}\frac{\partial ^2 g_{44}}{\partial\bar x^A
\partial\bar x^B}+\frac{1}{2}g^{EF}\,[AB,E]\,\frac{\partial g_{44}}{
\partial\bar x^F}-\frac{1}{4}{\rm e}^{2\psi -2\sigma}\frac{\partial g_{44}}
{\partial\bar z}\,\frac{\partial g_{AB}}{\partial\bar z}\ \nonumber\\
{} & {} & -\frac{1}{4}{\rm e}^{-2\psi}\,\frac{\partial g_{44}}{
\partial\bar x^A}\,\frac{\partial g_{44}}{\partial\bar x^B}\ .
\end{eqnarray}
Here $[AB,C]$ is the Christoffel symbol
\begin{equation}\label{B2}
[AB,C]=\frac{1}{2}\,\left (\frac{\partial g_{CA}}{\partial\bar x^B}
+\frac{\partial g_{BC}}{\partial\bar x^A}-\frac{\partial g_{AB}}{
\partial\bar x^C}\right )\ ,
\end{equation}
and
\begin{eqnarray}\label{B3}
{}^{(2)}\bar R_{ABCD} & = & \frac{1}{2}\,\left (g_{AD,BC}+
g_{BC,AD}-g_{AC,BD}-g_{BD,AC}\right ) \nonumber\\
{} & {} & +g^{EF}\,\left ([AD,E]\,[BC,F]-[AC,E]\,[BD,F]\right )\ ,
\end{eqnarray}
with the comma denoting partial derivative with respect to $\bar x^A$.

In the following two paragraphs we list some formulas which are 
required in section 4 above for the Aichelburg--Sexl boosts of $\bar R_{ijkl}$ 
in the $x$ and $z$ directions (i.e. transverse and parallel to the 
axis of symmetry respectively).

For a Lorentz boost in the $x$--direction, $\bar x^i\rightarrow x^i$ 
with
\begin{equation}\label{B4}
\bar x^1=\gamma\,(x^1-v\,x^4)\ ,\qquad \bar x^a=x^a\ ,\qquad 
\bar x^4=\gamma\,(x^4-v\,x^1)\ .
\end{equation}
Here $\gamma =(1-v^2)^{-1/2}$ and Latin letters $a, b, c, \dots$ 
take values $2,3$. Under this transformation the components $R_{ijkl}$ 
of the Riemann tensor in the unbarred coordinates are related to 
the barred components (\ref{B1}) by
\begin{eqnarray}\label{B5}
R_{abcd} & = & \bar R_{abcd}\ ,\qquad R_{ab14}=\bar R_{ab14}\ ,
\qquad R_{1414}=\bar R_{1414}\ ,\nonumber\\
R_{abc1} & = & \gamma\,\left (\bar R_{abc1}-v\,\bar R_{abc4}\right )\ ,\ \ 
R_{abc4}=\gamma\,\left (\bar R_{abc4}-v\,\bar R_{abc1}\right )\ ,
\nonumber\\
R_{a1b1} & = & \gamma ^2\left (\bar R_{a1b1}-v\,\bar R_{a1b4}-v\,\bar 
R_{a4b1}+v^2\bar R_{a4b4}\right )\ ,\nonumber\\
R_{a1b4} & = & \gamma ^2\left (\bar R_{a1b4}-v\,\bar R_{a1b1}-v\,\bar 
R_{a4b4}+v^2\bar R_{a4b1}\right )\ ,\nonumber\\
R_{a4b4} & = & \gamma ^2\left (\bar R_{a4b4}-v\,\bar R_{a1b4}-v\,\bar 
R_{a4b1}+v^2\bar R_{a1b1}\right )\ ,\nonumber\\
R_{a114} & = & \gamma\,\left (\bar R_{a114}-v\,\bar R_{a414}\right )\ ,
\ \ R_{a414}=\gamma\,\left (\bar R_{a414}+v\,\bar R_{a141}\right )\ .
\end{eqnarray}
When (\ref{B1}) is substituted into (\ref{B5}), the barred coordinates 
are expressed in terms of the unbarred coordinates using (\ref{B4}).

For a Lorentz boost in the $z$--direction $\bar x^i\rightarrow x^i$ 
with 
\begin{equation}\label{B6}
\bar x^A=x^A\ ,\qquad \bar x^3=\gamma\,(x^3-v\,x^4)\ ,\qquad 
\bar x^4=\gamma\,(x^4-v\,x^3)\ ,
\end{equation}
and $A=1,2$. The components $R_{ijkl}$ of the Riemann tensor in the 
unbarred coordinates are related to the barred components (\ref{B1}) by
\begin{eqnarray}\label{B7}
R_{ABCD} & = & \bar R_{ABCD}\ ,\qquad R_{AB34}=\bar R_{AB34}\ ,
\qquad R_{3434}=\bar R_{3434}\ ,\nonumber\\
R_{ABC3} & = & \gamma\,\left (\bar R_{ABC3}-v\,\bar R_{ABC4}\right )\ ,\ \ 
R_{ABC4}=\gamma\,\left (\bar R_{ABC4}-v\,\bar R_{ABC3}\right )\ ,
\nonumber\\
R_{A3B3} & = & \gamma ^2\left (\bar R_{A3B3}-v\,\bar R_{A3B4}-v\,\bar 
R_{A4B3}+v^2\bar R_{A4B4}\right )\ ,\nonumber\\
R_{A3B4} & = & \gamma ^2\left (\bar R_{A3B4}-v\,\bar R_{A3B3}-v\,\bar 
R_{A4B4}+v^2\bar R_{A4B3}\right )\ ,\nonumber\\
R_{A4B4} & = & \gamma ^2\left (\bar R_{A4B4}-v\,\bar R_{A3B4}-v\,\bar 
R_{A4B3}+v^2\bar R_{A3B3}\right )\ ,\nonumber\\
R_{A334} & = & \gamma\,\left (\bar R_{A334}-v\,\bar R_{A434}\right )\ ,
\ \ R_{A434}=\gamma\,\left (\bar R_{A434}+v\,\bar R_{A343}\right )\ .
\end{eqnarray}
When (\ref{B1}) is substituted into (\ref{B7}), the barred coordinates 
are expressed in terms of the unbarred coordinates using (\ref{B6}).

\setcounter{equation}{0}
\section{Boosting in the $-\bar z$ Direction}\indent
Starting with (\ref{4.16}) but now with $A_l=\gamma ^{-l-1}\,p_l$ 
for $l=0, 1, 2, \dots $, we arrive at $\psi$ given by (\ref{4.54}) 
with $\hat R , \hat w$ as in (\ref{4.55}) and (\ref{4.56}). The 
analogues of (4.19)--(4.21) are
\begin{eqnarray}\label{C1}
\frac{\partial\psi}{\partial\bar x} & = & -\sum_{l=0}^{\infty}
p_l\,x\,\frac{\gamma ^{-2l-4}}{\hat R^{l+3}}\,P'_{l+1}\ ,
\nonumber\\
\frac{\partial\psi}{\partial\bar y} & = & -\sum_{l=0}^{\infty}
p_l\,y\,\frac{\gamma ^{-2l-4}}{\hat R^{l+3}}\,P'_{l+1}\ ,
\\
\frac{\partial\psi}{\partial\bar z} & = & -\sum_{l=0}^{\infty}
p_l\,(l+1)\,\frac{\gamma ^{-2l-3}}{\hat R^{l+2}}\,P_{l+1}\ ,
\nonumber
\end{eqnarray}
where now the argument of the Legendre polynomials is $\hat w$ in 
(\ref{4.56}) and the prime denotes differentiation with respect to 
$\hat w$. When the relations (4.22), (4.23) satisfied by the Legendre 
polynomials are used again here with $w$ replaced by $\hat w$. We 
find that 
\begin{eqnarray}\label{C2}
\frac{\partial ^2\psi}{\partial\bar x^2} & = & -\sum_{l=0}
^{\infty}p_l\,\frac{\partial}{\partial x}\left (\frac{x\,\gamma ^{-2l-4}}
{\hat R^{l+3}}\,P'_{l+1}\right )\ ,\nonumber\\
\frac{\partial ^2\psi}{\partial\bar x\partial\bar y} & = & -\sum_{l=0}
^{\infty}p_l\,\frac{\partial}{\partial y}\left (\frac{x\,\gamma ^{-2l-4}}
{\hat R^{l+3}}\,P'_{l+1}\right )\ ,\nonumber\\
\frac{\partial ^2\psi}{\partial\bar y^2} & = & -\sum_{l=0}
^{\infty}p_l\,\frac{\partial}{\partial y}\left (\frac{y\,\gamma ^{-2l-4}}
{\hat R^{l+3}}\,P'_{l+1}\right )\ ,\nonumber\\
\frac{\partial ^2\psi}{\partial\bar z^2} & = & \sum_{l=0}^{\infty}
p_l\,(l+1)\,(l+2)\,\frac{\gamma ^{-2l-4}}{\hat R^{l+3}}\,P_{l+2}\ ,\\
\frac{\partial ^2\psi}{\partial\bar x\partial\bar z} & = & -\sum_{l=0}
^{\infty}p_l\,(l+1)\,\frac{\partial}{\partial x}
\left (\frac{\gamma ^{-2l-3}}
{\hat R^{l+2}}\,P_{l+1}\right )\ ,\nonumber\\
\frac{\partial ^2\psi}{\partial\bar y\partial\bar z} & = & -\sum_{l=0}
^{\infty}p_l\,(l+1)\,\frac{\partial}{\partial y}
\left (\frac{\gamma ^{-2l-3}}
{\hat R^{l+2}}\,P_{l+1}\right )\ .\nonumber
\end{eqnarray}
Limits involving these quantities as $v\rightarrow 1$ are required. 
In section 4 two results, (\ref{4.30}) and (\ref{4.35}), played a 
key role. A substantial difference in the present case is that 
these are replaced by the following:
\begin{eqnarray}\label{C.3,$}
\lim_{v\rightarrow 1}\frac{\gamma ^{-2}}{\hat R^{l+1}}\,P_l(\hat w) 
& = & 0\ ,\qquad l=0, 1, 2, \dots\ \ ,\\
\lim_{v\rightarrow 1}\frac{\gamma ^{-4}}{\hat R^{l+3}}\,P'_{l+1}(\hat w) 
& = & 0\ ,\qquad l=1, 2, 3,  \dots\ \ .
\end{eqnarray}
Both of these are established by induction on $l$. For (C.3) the 
result holds for $l=0$ since
\begin{equation}\label{C.5}
\lim_{v\rightarrow 1}\frac{\gamma ^{-2}}{\hat R}=0\ ,
\end{equation}
remembering that $\hat R$ is given by (\ref{4.55}). Assume (C.3) 
holds for $l$ and differentiate it with respect to $z$ to obtain
\begin{equation}\label{C.6}
\lim_{v\rightarrow 1}\frac{\gamma ^{-2}}{\hat R^{l+2}}\,
\left\{(l+1)\,\hat w\,P_l+(\hat w^2-1)\,P'_l\right\}=0\ .
\end{equation}
Now using (\ref{4.23}) with $w$ replaced by $\hat w$ this becomes 
\begin{equation}\label{C.7}
\lim_{v\rightarrow 1}\frac{\gamma ^{-2}}{\hat R^{l+2}}\,P_{l+1}=0\ ,
\end{equation}
showing that (C.3) holds for $l$ replaced by $l+1$. Thus on account 
of (\ref{C.5}) and (\ref{C.7}) we have established that (C.3) holds. 
A similar approach leads to (C.4). For $l=1$, (C.4) is true because 
\begin{equation}\label{C.8}
\lim_{v\rightarrow 1}\frac{\gamma ^{-4}}{\hat R^4}\,P'_2=3\,\lim_{v\rightarrow 1}
\frac{\gamma ^{-4}}{\hat R^5}\,(z-v\,t)=0\ .
\end{equation}
This last equality follows because we can write (\ref{3.5}) in the form
\begin{equation}\label{C.9}
\lim_{v\rightarrow 1}\frac{\gamma ^{-4}}{\hat R^5}=\frac{4}{3}\,
\frac{\delta (z-t)}{(x^2+y^2)^2}\ ,
\end{equation}
and $(z-t)\,\delta (z-t)=0$. Now assume (C.4) holds and differentiate 
it with respect to $z$ to find that
\begin{equation}\label{C.10}
\lim_{v\rightarrow 1}\frac{\gamma ^{-4}}{\hat R^{l+4}}\,\left\{
-(l+3)\,\hat w\,P'_{l+1}+(1-\hat w^2)\,P''_{l+1}\right\}=0\ .
\end{equation}
Using Legendre's differential equation for $P_{l+1}(\hat w)$ together 
with (\ref{4.22}) (with $w$ replaced by $\hat w$) we see that

\begin{equation}\label{C.11}
-(l+3)\,\hat w\,P'_{l+1}+(1-\hat w^2)\,P''_{l+1}=-(l+1)\,P'_{l+2}\ ,
\end{equation}
and so (\ref{C.10}) reads
\begin{equation}\label{C.12}
\lim_{v\rightarrow 1}\frac{\gamma ^{-4}}{\hat R^{l+4}}\,P'_{l+2}=0
\ .
\end{equation}
Using (\ref{C.8}) and (\ref{C.12}) we have established (C.4) by 
induction on $l$. We note that (C.4) also holds for $l=0$ but it 
is for $l\ge 1$ that we will now make use of it. The limits we shall 
need are the following: Using (C.3) we have
\begin{equation}\label{C.13}
\lim_{v\rightarrow 1}\gamma ^2\frac{\partial ^2\psi}{\partial\bar z^2}
=\sum_{l=0}^{\infty}p_l\,(l+1)\,(l+2)\,\gamma ^{-2l}\,
\left (\lim_{v\rightarrow 1}\frac{\gamma ^{-2}}{\hat R^{l+3}}\,P_{l+2}
\right )=0\ .
\end{equation}
Similarly
\begin{equation}\label{C.14}
\lim_{v\rightarrow 1}\gamma\,\frac{\partial ^2\psi}{\partial\bar x\partial
\bar z}=0\ .
\end{equation}
Next 
\begin{equation}\label{C.15}
\lim_{v\rightarrow 1}\gamma ^2\frac{\partial ^2\psi}{\partial\bar x^2}
=-\lim_{v\rightarrow 1}\sum_{l=0}^{\infty}p_l\,\frac{\partial}{\partial x}
\left (\frac{x\,\gamma ^{-2l-2}}{\hat R^{l+3}}\,P'_{l+1}\right )=
-\lim_{v\rightarrow 1}p_0\,\frac{\partial}{\partial x}\left (\frac{x\,\gamma ^{-2}}
{\hat R^{3}}\right )\ ,
\end{equation}
by (C.4) and since now (\ref{2.10}) is replaced by
\begin{equation}\label{C.16}
\lim_{v\rightarrow 1}\frac{\gamma ^{-2}}{\hat R^3}=\frac{2\,\delta (z-t)}
{x^2+y^2}\ ,
\end{equation}
we can write (\ref{C.15}) as
\begin{equation}\label{C.17}
\lim_{v\rightarrow 1}\gamma ^2\frac{\partial ^2\psi}{\partial\bar x^2}
=\frac{2\,p_0\,(x^2-y^2)}{(x^2+y^2)^2}\,\delta (z-t)\ .
\end{equation}
Similarly, using (C.4), we arrive at
\begin{eqnarray}\label{C.18}
\lim_{v\rightarrow 1}\gamma ^2\frac{\partial ^2\psi}
{\partial\bar y^2} & = & -\frac{2\,p_0\,(x^2-y^2)}
{(x^2+y^2)^2}\,\delta (z-t)\ ,\\
\lim_{v\rightarrow 1}\gamma ^2\frac{\partial ^2\psi}
{\partial\bar x\partial\bar y} & = & \frac{4\,p_0\,x\,y}
{(x^2+y^2)^2}\,\delta (z-t)\ .
\end{eqnarray}
As in the case of the boost in the $-\bar x$ direction described 
in section 4, the function $\sigma$ given by (4.3) does not 
survive in the limit $v\rightarrow 1$. Now $\bar R_{ijkl}$ 
is given by (B.1), $R_{ijkl}$ is given by (B.7) in the 
present case and from the latter we calculate $\tilde R_{ijkl}$ 
by taking the limit $v\rightarrow 1$. We find that the 
non--identically vanishing components are
\begin{eqnarray}\label{C.20}
\tilde R_{3232} &= & \tilde R_{2424}=-\tilde R_{2324}=-\tilde R_{3131}
=-\tilde R_{1414}=\tilde R_{1314} \nonumber\\
 & = & \lim_{v\rightarrow 1}2\,\gamma ^2
\frac{\partial ^2\psi}{\partial\bar y^2}=-\frac{4\,p_0\,(x^2-y^2)}
{(x^2+y^2)^2}\,\delta (z-t)\ ,
\end{eqnarray}
and
\begin{equation}\label{C.21}
\tilde R_{3231}=\tilde R_{2414}=-\tilde R_{1324}=-\tilde 
R_{2314}=\lim_{v\rightarrow 1}2\gamma ^2\frac{\partial ^2\psi}{\partial\bar x
\partial\bar y}=\frac{8\,p_0\,x\,y\,}{(x^2+y^2)^2}\,\delta (z-t)\ .
\end{equation}
This is essentially the same curvature tensor as in (3.7) and (3.8) 
in the Aichelburg--Sexl monopole case. With $p=-p_0$ and with $x$ 
and $z$ interchanged in (3.7) and (3.8) we obtain (C.20) and (C.21).

\end{document}